\documentclass[journal,10pt]{IEEEtran} 
\usepackage{cite}   
\usepackage[pdftex]{graphicx}   
\usepackage{array}  
\usepackage[font={small},labelfont={bf}]{caption}
\usepackage{subcaption}

\usepackage{amsmath,amssymb,amsfonts}  
\usepackage{textcomp}
\usepackage{xcolor} 
\usepackage{mathtools}
\usepackage{bm}    
\usepackage{comment}
\usepackage{booktabs}
\usepackage{algorithm,algorithmicx,algpseudocode}
\usepackage{multirow}
\usepackage{enumitem}

\graphicspath{{figs/}}  

\usepackage{stfloats} 

\usepackage{url}

\begin{document}
%
 \title{Joint Object Tracking and Intent Recognition}
%
%
%

\author{Jiaming~Liang,~
        Bashar~I.~Ahmad and Simon~Godsill,~\IEEEmembership{Fellow,~IEEE},
        
\thanks{J. Liang, B. I. Ahmad and S. Godsill are with the Department
of Engineering, University of Cambridge, Cambridge, CB2 1PZ, UK, e-mail: \{jl809, bia23, sjg30\} @cam.ac.uk.}
\thanks{This work was supported by a Defence and Security Accelerator DASA grant (DSTLX1000144447) and the SIGNetS project grant (W911NF17S0003-US-UK) by the US Army and UK Ministry of Defence.}


}

%
%

\markboth{}
{Liang \MakeLowercase{\textit{et al.}}: Bare Demo of IEEEtran.cls for IEEE Journals}
%
  
 

\maketitle

\begin{abstract}
This paper presents a Bayesian framework for inferring the posterior of the \textit{augmented} state of a target, incorporating its underlying goal or intent, such as any intermediate waypoints and/or the final destination. Thus, it is for joint object tracking and intent recognition. Several latent intent models are proposed here within a virtual leader formulation. They capture the influence of the target's hidden goal on its instantaneous behaviour. In this context, various motion models, including for highly maneuvering objects, are also considered. The \textit{a priori} unknown target intent (e.g. destination) can dynamically change over time and take any value within the state space (e.g. a location or spatial region). A sequential Monte Carlo (particle filtering) approach is introduced for the simultaneous estimation of the target's (kinematic) state and its intent. Rao-Blackwellisation is employed to enhance the statistical performance of the inference routine. Simulated data and real radar measurements are used to demonstrate the efficacy of the proposed techniques.
\end{abstract}


%
\IEEEpeerreviewmaketitle

%
%
%
%
\section{Introduction}
\IEEEPARstart{I}{n} recent years, there has been a growing interest in \textit{meta-level} tracking for achieving a higher-level understanding of complex scenes and its evolution over time by inferring and leveraging latent meta information on the observed targets, such as their intent and/or any social interactions within a group \cite{fanaswala2015spatiotemporal, ahmad18_bd,d2021malicious,liang2021SSPD, carravetta2021embedded, li2021sequential, tengesdal2022joint}. These can drive an object(s) long-term behaviour and future actions. The aim may be to circumvent conflict or identify opportunities, automate decision making, detect anomalies and optimize resources allocation (e.g. more effective operation of sensors and assets). Meta-level tracking has  numerous applications in surveillance, smart navigation, intelligent vehicles, cybernetics, to name a few.  In this paper, we tackle the problem of revealing, as early as possible, the target intent (namely destination and any trajectory waypoints) and accurately estimating its state $\mathbf{x}_t$ (e.g. position, velocity, higher order kinematics, etc.) from the available noisy sensory measurements. Conversely, a plethora of well-established \textit{sensor-level} tracking algorithms focus solely on determining $\mathbf{x}_t$,  the target's \textit{kinematic} state.

\subsection{Problem Statement}  \label{sec:problemstatement}
The objective here is to develop suitable stochastic models in conjunction with sequential inference methods to estimate the target's intent and kinematic state such that one or more of the following can apply:
\begin{itemize}
    \item The intent (e.g. destination) can dynamically change over time and take any value within the state space (e.g. a location or region within a geographical area constituting a sensor field of coverage).
    \item The object trajectory can contain a set of waypoints, each can be treated as an intermediate (or quasi) destination. This is a common behaviour of (semi-)autonomous systems such as drones \cite{beard2012small} and even in manned aviation or maritime navigation, for example an airliner or a vessel follows a prescribed route with waypoints \cite{seah2009stochastic, millefiori2016modeling, rezaie2021conditionally}.
    \item The target can be agile (e.g. a rotary wing small unmanned air system UAS or drone) and capable of undertaking sharp sudden as well as frequent manoeuvres such as swift turns or abrupt accelerations.
\end{itemize} 

\subsection{Related Work}\label{sec:relatedwork}
Exploiting known constraints on an object's latent state (e.g. future values such as final destination, possible routes, etc.) is widely recognised to enhance the quality of the state estimation. This can be attributed to the original work by \textit{Castanon et. al.} \cite{castanon1985algorithms}. Several relevant modelling approaches have been introduced since that incorporate such constraints, for example with reciprocal processes \cite{baccarelli1998recursive, carravetta2012modelling, white2020state, fanaswalaJSTSP2013, rezaie2019gaussian}, conditionally Markov (CM) processes \cite{rezaie2019gaussian, arezaie2020destination, rezaie2021conditionally, lee2024destination}, embedded stochastic syntactic processes  \cite{carravetta2021embedded}, context-free grammars \cite{fanaswalaJSTSP2013, fanaswala2015spatiotemporal, Krishnamurthy2018}, pseudo-measurements with traditional Markov models \cite{tahk1990target, liang19_pseudo, zhou2018pseudo, zhou2020},  Ornstein–Uhlenbeck-type processes \cite{millefiori2016modeling, ahmad16_ou, liang19_pseudo, tengesdal2022joint}, intrinsic coordinate model \cite{liang2020MLSP} and by embedding the endpoint information into an (unconstrained) motion dynamics \cite{xu2022modeling,linfeng2024}.

Unlike the majority of the above destination-constrained models, inclusive of those specifically devised for waypoints-driven trajectories \cite{seah2009stochastic, rezaie2021conditionally}, in this paper the \textit{latent} intent, which can change over time, is modeled as a continuous-time stochastic process with its own dynamical model. It is then linked to the target kinematic state, where both form an \textit{augmented} state, within a Bayesian framework using the virtual leader modeling approach. For example, in \cite{rezaie2021conditionally} destination/waypoint is modeled by a random variable with a variance as prior information, implying fixed intent. Nonetheless, in \cite{brezaie2020mathematical} a time-varying destination or guide is modeled by a discrete stochastic process based on a class of CM sequences. Here, we combine target kinematic state and intent models in a continuous-time space under the common Markov assumption, facilitating applying established inference routines for joint tracking and intent prediction. Compared to discrete models, this can be advantageous in practice, such as for handling different and varying update rates for the target intent, kinematic state and sensory measurements.

The proposed formulation in this paper is motivated by the virtual leader  \cite{mahler2007book,skpang2011} and bulk \cite{salmond99} models, which have proved effective in tracking groups of objects \cite{skpang2011}. The idea is to model the target intent (e.g. final destination and/or waypoints) as a virtual leader (VL)  that has its own dynamics. This serves the purpose of connecting the object motion to its intent through carefully designed stochastic equations. The influence of the target's medium to long term goal on its instantaneous behaviour is thus captured. This is fundamentally distinct from the related work on intent prediction in \cite{ahmad18_bd, liang19_pseudo,liang2020MLSP,lyudmil20,AhmadMag2017, gan2019bayesian, tengesdal2022joint}, where the target destination belongs to a predefined finite set of nominal endpoints at fixed locations/values; see \cite{gan_liang_ahmad_godsill_2020, gan2021levy, tengesdal2022joint} for an overview. Owing to the VL modelling, the target intent here can take any value within the state space and it can dynamically change over time, independently from the sensory observations.

In this paper, we build on our preliminary work in \cite{liang2021SSPD} and \cite{gan2021levy}, which only discusses the rudimentary VL model of the target final destination in Section \ref{ssec:vl_basic_des_model}. 
For example, in \cite{liang2021SSPD} only linear Gaussian models are treated. Below, we introduce various latent intent models, including for waypoints-driven trajectories with a piecewise constant model with jumps. This can be combined  with jump diffusion motion models \cite{christensen2012forecasting, gan2019bayesian} for fast maneuvering targets
. The overall proposed modelling here leads to a jump particle filter \cite{godsill07,bunch2013vr} for estimating the sought joint posterior of the target kinematic state and intent. 

Context-free grammars (CFG) and other natural language processing models, with a discrete state space, are utilised in \cite{fanaswalaJSTSP2013,fanaswala2015spatiotemporal,Krishnamurthy2018} for meta-level tracking, namely to detect the target intent and anomalous behavior. For example, CFGs stipulate that an object should pass through a finite number of specified grid cells to reach an endpoint via a viable path. Furthermore, the hybrid modeling with path-constrained dynamic optimization in \cite{fu2025inner} incorporates environmental constraints (e.g. roads) to infer a target endpoint. A pseudo-measurement technique for estimating the target state and destination is also described in \cite{zhou2020}. It is based on a linear equality constraint such that the object is assumed to follow a simple path (e.g. a straight line) to its intended endpoint. On the other hand, we recall that the approach introduced in this work employs continuous state space models, where targets can follow complex trajectories. This permits treating asynchronous sensory measurements and does not restrict the target motion behavior unlike \cite{fanaswala2015spatiotemporal ,Krishnamurthy2018, zhou2020, fu2025inner}.

A related modeling problem is that for pursuit and evasion scenarios such as predator-prey interactions amongst animals, dogfighting in aerial combat and missile pursuit with target evasion \cite{pais2010pursuit}. Whilst the presented modeling and other cited related work can be adapted for such settings, the motion of the pursued entity and its evasion tactics are not implicitly modeled here. This can be considered in future work.

Finally, there are several data-driven techniques, e.g. using deep neural networks, physics-informed networks or Gaussian processes, for intent modelling and prediction \cite{gaurav2019discriminatively, radwan2020multimodal,rudenko2020human, xu2022intention, perrusquia2024uncovering, perrusquia2024DMD, ZHAO2025}. They typically rely on previously observed data to learn the target motion model and forecast its future locations. This is often applied when the considered scenario is simple and/or object movements are confined to within small easily observed areas (e.g. a communal space or road junction or pedestrian crossing with limited set of exits or routes for pedestrians and cars) and/or (partial) access to target motion control signals. Here, we propose a model-based approach within an object tracking setting. It relies on appropriately designed stochastic models of the object latent state and intent. As is common in object tracking, these models have a few unknown parameters which are physically meaningful and can be set manually or learnt from data (e.g. based on the maximum likelihood criterion).

\subsection{Contributions}\label{sec:contributions}
The main contributions of this paper are:
\begin{itemize}
    \item Novel latent intent models based on the virtual-leader formulation such that the target kinematic state and its unknown intent are intrinsically linked to capture, in continuous-time, the influence of the latter on the former. In particular, the continuous and piecewise-constant model represents slow and/or sudden changes in the destination or waypoint.  Fast manoeuvring objects can be additionally treated within the same framework. 
    \item A Rao-Blackwellised variable rate particle filtering approach to sequentially infer the joint posterior of the target latent kinematic state and intent from noisy and potentially asynchronous sensory observations. It offers additional flexibility in terms of permitting spontaneously examining the likelihood that any specific spatial location or extended area is the destination of the tracked object.  
\end{itemize}
Results from synthetic and real radar data are shown to demonstrate the performance of the proposed methods. 
\subsection{Paper Outline}\label{sec:paperoutline}
The remainder of the paper is organised as follows. Several models are introduced in Section \ref{sec:ProposedModels} and a particle-filtering-based inference algorithm is presented in Section \ref{sec:inference_method}. Results from synthetic and real data are shown in \ref{sec:Results}. Conclusions are drawn in Section \ref{sec:conclusion}. 
  
\section{Latent Intent Models} 
\label{sec:ProposedModels}

In this section, we introduce a range of novel models, each of which represent different target and intent dynamics. This is preceded by describing the basic, baseline, model in \cite{liang2021SSPD} for completeness. 
Without loss of generality, the unobserved target intent, denoted as $\mathbf{r}$, will be considered to be the Cartesian coordinates of the intended endpoint of a target in the remainder of the paper. 

It should be emphasised that the latent intent variable below can be straightforwardly extended to model higher order kinematics (e.g. velocity and acceleration). The impact of including \textit{informative} priors on such higher order kinematics can improve the intent inference accuracy; otherwise their impact on the destination prediction can be marginal. For example, in human machine interaction applications the inclusion of the velocity of a tracked freehand pointing gesture/finger at the intended destination can be based on collected experimental interactions data as in \cite{AhmadMag2017, gan2020modeling}.  A velocity and acceleration profiles of a (semi-)autonomous platform, such as a drone,  at/near the intended destination is dependent on its capabilities, role and mission (e.g. nearly constant cruising speed during surveillance for fixed-wing UAS or gradual reduction to hovering for rotary-wing versus high speed for loitering munitions).

\subsection{Baseline Model}
\label{ssec:vl_basic_des_model}
Here, both the target motion and the destination are driven by Brownian motions. Consider first a 1-dimensional motion in Cartesian coordinate and denote the overall (i.e. augmented) state vector at time $t$ as $\mathbf{s}_t = [ \mathbf{x}_t, \mathbf{r}_t]^{T}$ with $\mathbf{r}_t=r^x_t$ being the endpoint location along the $x$-axis and $\mathbf{x}_t=[x_t, \dot{x}_t]^T$ the target position and velocity. The virtual-leader based latent intent/destination model can now be described by the following stochastic differential equations (SDEs) \cite{oksendal2003stochastic}, 
\begin{align} 
    d \dot{x}_t & = \eta_x ( r^x_t - x_t)dt - \rho_x \dot{x}_t dt + \sigma_x d B_t, \label{eqn:vl_basic_vel} \\
    d r^x_t & = \sigma_r dB_t,   \label{eqn:vl_basic_des}
\end{align} 
with $\eta_x$, $\rho_x$ being the mean reversion constants (specifically $\rho_x$ is the drag coefficient, preventing the corresponding velocity from drifting to large values over time) and $\sigma_x$, $\sigma_r$ the diffusion constants. $B_t$ is a 1-dimensional standard Brownian motion. Equation \eqref{eqn:vl_basic_vel} depicts a destination-reverting behaviour of the target position, hence leading to a destination-driven target dynamics. Its second component serves to prevent the velocity from being excessively high. 

The target intent is allowed to change dynamically with $\sigma_r \neq 0$ in \eqref{eqn:vl_basic_des} or it can be fixed with $\sigma_r=0$ (i.e. in the sense of having the minimum uncertainty about the destination). This model can be formalised in the vector-matrix form
\begin{align} \label{eqn:vl_basic_linearSDE}
    d \mathbf{s}_t = A \mathbf{s}_t dt + L dB^s_t,
\end{align}
with
\begin{align}
    A =
    \begin{bmatrix}
    0 & 1 & 0 \\
    -\eta_x & -\rho_x & \eta_x \\
    0 & 0 & 0
    \end{bmatrix},
    L = 
    \begin{bmatrix}
    0 & 0 \\
    \sigma_x & 0 \\
    0 & \sigma_r
    \end{bmatrix},
    B^s_t = [ B_t, B_t]^T.
    \label{eqn:vl_basic_des_matrices}
\end{align}
A system model in a $k$-dimensional Cartesian coordinate system can be obtained readily by cascading several such 1-dimensional models. It can be shown that the solution to the linear time-invariant SDE \eqref{eqn:vl_basic_linearSDE} is given by:
\begin{equation} \label{eqn:vl_basic_state_transition_integral}
    \mathbf{s}_n = e^{A\tau} \mathbf{s}_{n-1} +   \int^{\tau}_0 e^{A (\tau-u ) } L dB^s_{u},  
\end{equation}
with $\tau = t_n - t_{n-1}$ being the time step between two successive time instants and $e^{(\cdot)}$ is the matrix exponential operator. Correspondingly, the state transition density is given by
\begin{equation} \label{eqn:vl_linear_gauss_density}
    p(\mathbf{s}_n|\mathbf{s}_{n-1}) = \mathcal{N}(\mathbf{s}_{n} \vert F_\tau \mathbf{s}_{n-1}, Q_\tau  ),
\end{equation}
where
\begin{align}  \label{eqn:expm(A)_basic}
    F_{\tau} & = e^{A\tau}, ~~~
    Q_{\tau} & = \int^{\tau}_0 e^{A (\tau-u ) } LL^T {e^{A (\tau-u ) }}^T du,
\end{align}
and $\mathcal{N}(\cdot | \cdot)$ represents the normal distribution. The state process covariance $Q_{\tau}$ can be obtained in a closed form or via using the matrix fraction decomposition \cite{sarkka_solin_2019} to simplify the calculation (especially in high-dimensional cases). Note that higher-order kinematics may be augmented to the latent intent variable, e.g. velocity at endpoint $\dot{r}^x_t$ and $\mathbf{r}_t = [r^x_t, \dot{r}^x_t]^T$. In this case, it resembles the group virtual leader model in \cite{skpang2011}, albeit with the number of objects equal to one. 

This rudimentary intent-driven model of linear and Gaussian structure, for example see \cite{liang2021SSPD} and \cite{gan2021levy}, serves here as a baseline method for performance benchmarking in Section \ref{sec:Results}.
 
\subsection{Piecewise-constant Destination Model} \label{ssec:vl_piecewiseconst}
The above basic model implicitly stipulates that the target destination evolves over time in synchronisation with the sensor(s) measurements. This limits its applicability since in many practical situations the intent is expected to change at a significantly slower rate compared with the sampling frequency of the sensor measurements. 
For instance, a (semi-)autonomous system, such as a drone, following a way-point trajectory set during the mission planning, will typically have one or more ``intermediate`` destinations (i.e. waypoints) between which the intent remains unchanged. This is despite potentially many sensor observations (e.g. from radar or optical sensors) made during this period. Hence, to better characterise a target motion with \textit{a priori} unknown waypoints or sojourns, we use the same object (kinematic) state model as in (\ref{eqn:vl_basic_linearSDE}) with $t \in (\tau_k, \tau_{k+1}]$. The latent destination $\mathbf{r}_k = r_k^x$ is assumed to remain unchanged within the time interval $(\tau_k, \tau_{k+1}]$ and it has a distinctive subscript (i.e. $k$) compared with that in \eqref{eqn:vl_basic_des}. Consequently, the target position is driven stochastically towards some  destination $r^x_k$ before the arrival of an updated destination $r^x_{k+1}$ right after $\tau_{k+1}$.

Consequently, the latent destination follows a continuous-time piecewise deterministic process (PDP) \cite{Davis1984PDP} and its value only changes randomly at a collection of countable jumping times $\{ \tau_k \}$. In principle, the intent can be drawn from any suitably chosen density. Here and for simplicity, we model it as a jump process $W_t$ with independent and normally distributed jump sizes \{$J_k$\} since this can lead to a conditionally linear and Gaussian structure for the dynamical model. We show in Section \ref{sec:inference_method} that this particular structure facilitates the use of the Rao-Blackwellised estimation scheme that can improve the efficiency of the standard particle filter when the measurement model is linear-Gaussian. Specifically, we have
\begin{align} \label{eqn: vl_des_jump_component}
    d r^x_t = dW_t,
\end{align} 
\begin{align*} 
    dW_t &= 
    \begin{cases}
      J_k, & t=\tau_k \\ 
      0, & \text{elsewhere}
    \end{cases}  \\
    J_k & \sim \mathcal{N}(\mu_J, \sigma_J^2).
\end{align*}   

We recall that the \textit{a prior} unknown intent can take any value within the object's state space (i.e. modeled by a continuous-time process). Consequently, the  approach in \eqref{eqn: vl_des_jump_component} suitably models potential abrupt changes in the object intent, which is reflected in the object (kinematic) state evolution over time aimed at achieving a newly set goal (e.g. reaching a destination/waypoint in the scene). Duration between successive changes in intent (i.e. sought destination/waypoint) can be assumed to be sufficiently large, compared to the target kinematic state update rate. Hence, it is reasonable for the corresponding jumps to be assumed independent. Alternatively, jump process with correlated jump times and sizes can be used within the presented modeling and particle filtering framework \cite{godsill07esaim}. This requires defining corresponding priors that encapsulate such dependencies, e.g. as in financial modeling applications \cite{cont2003financial}.   

The SDE for the 1-dimensional state vector $\mathbf{s}_t = [x_t,\dot{x}_t,r_t^x]^T$ can then be written as
\begin{equation} \label{eqn:vl_piecewiseconst_SDE}
    d \mathbf{s}_t = A \mathbf{s}_t dt + \mathbf{h}_J dW_t + \mathbf{h}_B dB_t,
\end{equation}
with 
\begin{align*} 
    \mathbf{h}_J = 
    \begin{bmatrix}
    0 & 0 & 1
    \end{bmatrix}^T,
    \mathbf{h}_B = 
    \begin{bmatrix}
    0 & \sigma_x & 0 
    \end{bmatrix}^T.
\end{align*} 
 \
 
 The resulting model can be seen as a variable rate model \cite{godsill07,bunch2013vr} since the changepoint pairs $\{ J_k, \tau_k \}$ are decoupled from the observation process. For more details regarding variable rate models, see \cite{godsill07,Whiteley2011PDP,bunch2013vr}. If the random jump times $\{\tau_k \}_{k:\tau_k \in [t_{n-1}, t_n) }$ are assumed known, the following conditionally linear-Gaussian state transition density can be obtained by solving \eqref{eqn:vl_piecewiseconst_SDE} as in \cite{godsill07esaim},
\begin{align} \label{eqn:vl_conditional_pc_density}
    p(\mathbf{s}_n|\mathbf{s}_{n-1}, \{\tau_k \}_{k:\tau_k \in (t_{n-1}, t_n] }) = \mathcal{N} (\mathbf{\mu}_{n:n-1}, \Sigma_{n:n-1}),
\end{align}
where
\begin{align*}
    & \mathbf{\mu}_{n:n-1} = e^{A(t_n-t_{n-1})}\textbf{s}_{n-1} + \sum_{k: \tau_k \in [t_{n-1}, t_n)} \mu_J  e^{A(t_n-\tau_k)} \mathbf{h}_J,  \\
    & \Sigma_{n:n-1} =  \sigma_J^2 \sum_{k: \tau_k \in [t_{n-1}, t_n)} e^{A(t_n-\tau_k)} \mathbf{h}_J \mathbf{h}_J^T {e^{A(t_n-\tau_k)}}^T \\
    & \;\; +  \int^{t_n-t_{n-1}}_0 e^{A(t_n-t_{n-1}-u)} \mathbf{h}_B \mathbf{h}_B^T {e^{A(t_n-t_{n-1}-u)}}^T du.
\end{align*} 

The inter-arrival times between jumps are assumed to follow an appropriately chosen distribution. As an example and for simplicity, we utilise in this paper the Gamma distribution,
\begin{equation} \label{eqn:shifted_gamma}
    \tau_k-\tau_{k-1} \sim \mathrm{Gam}(\alpha_{\tau}, \beta_{\tau}),
\end{equation}
where $\alpha_{\tau}$, $\beta_{\tau}$ are the shape and scale parameters, respectively. Besides its convenient statistical properties (including conjugacy), a Gamma distribution is flexible and adaptable which facilitates modeling different behaviors and levels of skewness via adjusting its two parameters. We emphasise that any other suitable distribution can be adopted, e.g. the exponential distribution  in \cite{christensen2012forecasting} which is a special case of (\ref{eqn:shifted_gamma}).

This model can maintain a consistent intent for a period of time determined by two consecutive jump times. Beyond way-point driven trajectories, it can be applied with the velocity-reverting models in \cite{millefiori2016modeling, tengesdal2022joint} for tracking and anomaly detection in maritime surveillance where prescribed navigation routes can be represented by a set of piecewise constant typical velocity profiles. Additionally, we present in Appendix A the \textit{multi-hypothesis intent model}, which is based on this piecewise constant destination one. It is specifically for examining a finite set of predefined $N_{\mathcal{D}}$ destinations. However, unlike prior related work, e.g. \cite{ahmad18_bd, gan_liang_ahmad_godsill_2020, liang19_pseudo,liang2020MLSP, AhmadMag2017, gan2019bayesian, gan_liang_ahmad_godsill_2020, gan2021levy}, it explicitly incorporates the null-hypothesis, i.e. none of the prescribed endpoints are sought by the target.  
\subsection{Jump-diffusion Destination Model} \label{ssec:vl_jumpdiff}
Following from the previous piecewise constant formulation and as a generalisation, the target latent destination can evolve according to the following jump-diffusion model
\begin{align} \label{eqn:vl_jump_diffusion}
    d r^x_t  = dW_t +  \sigma_{r} dB_t,
\end{align}  
to accommodate both minor and major changes in the underlying target intent. The resultant SDE for $\mathbf{s}_t$ is
\begin{equation} \label{eqn:vl_JPDM_SDE}
    d \mathbf{s}_t = A \mathbf{s}_t dt + \mathbf{h}_J dW_t + LdB^s_t,
\end{equation}
which follows from \eqref{eqn:vl_piecewiseconst_SDE} with $\mathbf{h}_B dB_t$ replaced by $LdB^s_t$ as in \eqref{eqn:vl_basic_linearSDE}. The piecewise constant model in \eqref{eqn:vl_piecewiseconst_SDE} can thus be obtained from \eqref{eqn:vl_jump_diffusion} by setting $\sigma_{r}=0$.

The destination dynamics here can be viewed as a combination of the two behaviours:
\begin{enumerate}[label=\alph*)]
    \item \textit{exploration:} controlled by $W_t$, with which large step sizes can be taken within the state space when there is a drastic change in intent, i.e. macro intent-driven motion; and
    \item \textit{exploitation:} governed by $B_t$, with a relatively small diffusion constant $\sigma_{r}$, and gives the model the ability to make smaller adjustments (e.g. finer corrections) in the vicinity of the endpoint, i.e. micro intent-driven motion.
\end{enumerate}
This model is suitable for scenarios where intermediary waypoints and final destinations are briskly chosen and refined during the target journey, in lieu of being set in advance. For instance, a manually controlled UAV (e.g. with a first-person view FPV camera) can suddenly identify and home in on a location of interest during the flight. 

\subsection{Fast Manoeuvring Target with Latent Destination} \label{ssec:vl_fmt}
Here, we extend the modelling to represent fast manoeuvring behaviour that can entail the target undertaking swift turns and/or accelerations, for instance due to sudden change in intent, obstacle avoidance, and being subjected to perturbations (e.g. induced by external environmental factors such as wind, vibrations, turbulence, etc.). The object motion model along $x$ axis in this case is described by the SDE
\begin{align} \label{eqn:jump_diff_target}
    d \dot{x}_t & = \eta_x ( r^x_k - x_t)dt - \rho_x \dot{x}_t dt + \sigma_x d B_t + dW_t, 
\end{align}  
with the introduction of the jump component $dW_t$ into \eqref{eqn:vl_basic_vel}. For simplicity, this is combined with the (baseline) Brownian motion-driven latent destination model in \eqref{eqn:vl_basic_des}. The SDE for the state vector $\mathbf{s}_t=[x_t, \dot{x}_t, r^x_t]^T$ in this case is given by
\begin{equation} \label{eqn:target_jump_diff_gauss_leader_SDE}
    d \mathbf{s}_t = A \mathbf{s}_t dt + \mathbf{h}_J dW_t + L dB^s_t,
\end{equation}
with $\mathbf{h}_J$, $A$ and $L$ defined in \eqref{eqn:vl_basic_des_matrices}.

This model can be solved in exactly the same manner as the piecewise constant and the jump-diffusion destination models because they share the same parametric form. The distinction is that the destination variable will neither jump nor remain piecewise deterministic as it is now driven only by Brownian motion. Nonetheless, target motion and intent models each containing a separate jump process, i.e. \eqref{eqn:vl_jump_diffusion} and \eqref{eqn:jump_diff_target}, with de-coupled jump times, can also be considered within the presented modelling framework; however this is outside the scope of this paper. Furthermore, it is worth noting that the target dynamical model is not limited to the jump-diffusion model, any other models suitable for fast manoeuvring target may be employed to  in place of that in \eqref{eqn:jump_diff_target}-\eqref{eqn:target_jump_diff_gauss_leader_SDE}, e.g. the stable L\'{e}vy state-space model as in \cite{gan2021levy}. 

\section{Inference Methodology} \label{sec:inference_method}

\subsection{System Model and Inference Task}
With the dynamical models, we have the observation model
\begin{equation*}
\mathbf{m}_n \sim p( \mathbf{m}_n | \mathbf{s}_{n}),
\end{equation*}
where $p(\cdot)$ is some known distribution characterising the observation process and the noisy measurements $\mathbf{m}_{0:n}=\{ \mathbf{m}_0,\mathbf{m}_1,\ldots,\mathbf{m}_n \}$ made at time instants $t_{0:n}=\{ t_0,t_1,\ldots,t_n \}$; $t_{0:n}$ need not be regularly spaced and asynchronously sampled data can be naturally treated under the adopted continuous-time state-space modelling. For simplicity, we assume below the linear Gaussian observation model
\begin{equation} \label{eqn:linear_gauss_measurement_model}
    \mathbf{m}_n \sim p( \mathbf{m}_n | \mathbf{s}_n)=\mathcal{N}(\mathbf{m}_n| H_n \mathbf{s}_n, R_n),
\end{equation}
where $R_n$ is the noise covariance matrix encoding the sensory data uncertainty. The mapping matrix $H_n$ extracts the entries of the overall state vector $\textbf{s}_n$ (e.g. Cartesian coordinates of the object location), except those associated with the unobservable latent intent $\mathbf{r}_n$. Fixed matrices are assumed here, i.e. $H_n=H$ and $R_n=R$. Nevertheless, the models in Section \ref{sec:ProposedModels} can be used in conjunction with any other measurement models, including non-linear and/or non-Gaussian ones \cite{doucet2000sequential}.  

We recall that the objective of the tackled joint intent and kinematic state estimation task is to sequentially estimate the posterior distribution $p(\mathbf{s}_n|\mathbf{m}_{0:n})$ or $p( \mathbf{s}_n, \bm{\theta}_{t_0:t_n})$ at each time step $t_n$ from all of the available data $\mathbf{m}_{0:n}$. Whilst the baseline model in Section \ref{ssec:vl_basic_des_model} with \eqref{eqn:linear_gauss_measurement_model} implies that the standard linear Kalman filter is the optimal estimator, the proposed variable rate intent-driven models in this paper do not lead to an analytically tractable solution for inferring the sought posterior distribution. This is due to the inclusion of a change-point sequence. We hence present in this paper a jump particle filtering approach, termed variable rate particle filter (VRPF) \cite{godsill07,bunch2013vr}. For convenience, we use  $\bm{\theta}_k= \tau_k $ to denote the variable rate state with $\bm{\theta}_{t_0:t_n}=\{ \bm{\theta}_k\}_{k: t_0 \leq \bm{\theta}_k < t_n}$ as the collection of variable rate state points between $t_0$ and $t_n$. Hence we first specify the variable rate state priors before giving a detailed description of the state inference algorithm.
  
\subsection{Variable Rate State Priors}
\label{ssec:vr_priors}
From \cite{jacobsen2006,Whiteley2011PDP,bunch2013vr}, the prior for $\bm{\theta}$ can be expressed as
\begin{align}\label{eqn:theta_prior}
    p(\bm{\theta}_{t_0:t_n}) = S(t_n, \tau_{K_n}) \prod_{k=1}^{K_n} p(\tau_k|\tau_{k-1}),
\end{align}
where $S(t_n,\tau_k)=\int^{\infty}_{t_n} p(u|\tau_k)du $ is the conditional probability of no additional jump occurring before time instant $t_n$, $K_n = \max \{k: \tau_k < t_n\}$ and by convention $\tau_0=t_0=0$; refer to Appendix A for the corresponding definition of $\mathbf{\theta}_n$ and $p(\bm{\theta}_{t_0:t_n})$ of the multi-hypothesis model.

Similar to standard filtering methods where the inference is normally performed at the observation times, we formulate the VRPF in terms of $\mathbf{t}_{0:n}$ albeit the jumps time are not necessarily synchronised with those of measurements. To this end, it is necessary to obtain the variable rate state transition density from $t_{n-1}$ to $t_n$ using the joint priors, 
\begin{align} \label{eqn:variable_rate_state_transition_density}
    & p( \bm{\theta}_{t_{n-1}:t_n} | \bm{\theta}_{t_0:t_{n-1}} ) 
     = \frac{ p(\bm{\theta}_{t_0:t_n}) }{ p(\bm{\theta}_{t_0:t_{n-1}}) } \notag \\
    & \qquad \qquad = \frac{S(t_n, \tau_{K_n})}{S(t_{n-1}, \tau_{K_{n-1}})} \prod_{k: t_{n-1} \leq \tau_k < t_n} p(\bm{\theta}_k | \bm{\theta}_{k-1}).
\end{align}
A pragmatic way to propose state sequence $ \bm{\theta}_{t_{n-1}:t_n}$ from  \eqref{eqn:variable_rate_state_transition_density} is to draw $\bm{\theta}$'s until the most recently simulated jump time falls beyond $t_n$ and then keep the samples falling within $[t_{n-1}, t_n)$.

\subsection{Rao-Blackwellised Variable Rate Particle Filter}
The target posterior distribution at $t_n$ can be factorised as
\begin{align} \label{eqn:rbvrpf_target_distribution_factorization}
    p( \mathbf{s}_n, \bm{\theta}_{t_0:t_n} | \mathbf{m}_{0:n}) = p( \mathbf{s}_n | \bm{\theta}_{t_0:t_n}, \mathbf{m}_{0:n})p(\bm{\theta}_{t_0:t_n}|\mathbf{m}_{0:n}).
\end{align}
From Section \ref{sec:ProposedModels} the state transition densities are linear and Gaussian conditional upon the jump time sequence (as well as the indicator sequences $\{ c \}$ for the multi-hypothesis model) and the measurement model is linear-Gaussian. This implies that tractable filtering for the (augmented) system state, namely the estimation of the multivariate Gaussian $p( \mathbf{s}_n | \bm{\theta}_{t_0:t_n}, \mathbf{m}_{0:n})$, can be performed in an optimal manner via the Kalman filter. The variable rate particle filter then needs only to operate on the marginal distribution $p(\bm{\theta}_{t_0:t_n}|\mathbf{m}_{0:n})$. This inference procedure can lead to a reduction in the variance of the estimates and it is commonly known as Rao-Blackwellisation \cite{chen2000mixture, Robert2004}. The VRPF in this case is termed Rao-Blackwellised VRPF (RBVRPF) \cite{godsill07esaim,morelande2009,christensen2012forecasting}, which is analogous to the standard Rao-Blackwellised particle filter \cite{RBPF_liu_chen_1998,doucet2000sequential,schon05}. 

 
We now detail the particle filtering approach which estimates the posterior distributions of the variable rate state sequence $\bm{\theta}_{t_0:t_n}$ and the linear state $\mathbf{s}_n$ at time instant $t_n$. Similar to the standard Rao-Blackwellised particle filter, at $t_{n}$ the RBVRPF maintains a collection of $N_p$ weighted particles 
\begin{align*}
    \{ \omega^{(i)}_n, \bm{\theta}_{t_0:t_n}^{(i)}, \mu_{n|n}^{(i)}, \Sigma_{n|n}^{(i)} \}_{1 \leq i \leq N_p},
\end{align*} 
where $ \omega_n^{(i)}$ is the normalised particle weight associated with the $i$-th particle and $\sum^{N_p}_{i=1} \omega_n^{(i)} = 1$. Given the particle set, the posterior density for $\theta_{t_0:t_n}$ can be approximated as
\begin{align} \label{eqn:vrpf_posterior_delta}
    {p}(\bm{\theta}_{t_0:t_n}|\mathbf{m}_{0:n}) \approx \sum^{N_p}_{i=1} \omega^{(i)}_n \delta_{\bm{\theta}_{t_0:t_n}^{(i)}} (\bm{\theta}_{t_0:t_n})  .
\end{align}
Each particle also stores the mean and the covariance parameters of the following Gaussian distribution 
\begin{align} \label{eqn:kalman_posterior_part_i}
    p(\mathbf{s}_n | \bm{\theta}_{t_0:t_n}^{(i)}, \mathbf{m}_{0:n}) = \mathcal{N} (\mathbf{s}_n | \mu_{n|n}^{(i)}, \Sigma_{n|n}^{(i)} ),  
\end{align}
which is computed by a Kalman filter running on the $i$-th particle. Substituting \eqref{eqn:kalman_posterior_part_i} and \eqref{eqn:vrpf_posterior_delta} into \eqref{eqn:rbvrpf_target_distribution_factorization} and marginalising over the variable rate state sequence, the posterior density for the linear state is approximated by a mixture of Gaussians
\begin{align} \label{eqn:linear_posterior_gauss_mixture}
    p(\mathbf{s}_n | \mathbf{m}_{0:n}) & = \int p( \mathbf{s}_n | \bm{\theta}_{t_0:t_n}, \mathbf{m}_{0:n})p(\bm{\theta}_{t_0:t_n}|\mathbf{m}_{0:n}) d \bm{\theta}_{t_0:t_n}  \notag \\
    & \approx \sum^{N_p}_{i=1} \omega^{(i)}_n \mathcal{N} (\mathbf{s}_n | \mu_{n|n}^{(i)}, \Sigma_{n|n}^{(i)} ).
\end{align}
Calculating \eqref{eqn:vrpf_posterior_delta} and \eqref{eqn:linear_posterior_gauss_mixture} entails the following two operations.

\subsubsection{Compute particle weights}
Assuming that a weighted set of particles has been obtained from the last time instant $t_{n-1}$, the target posterior distribution of the particle filter at $t_n$ is 
\begin{align} \label{eqn:target_density_rbvrpf}
    p(\bm{\theta}_{t_0:t_n}|\mathbf{m}_{0:n}) 
    & \propto p(\mathbf{m}_n | \bm{\theta}_{t_0:t_{n}}, \mathbf{m}_{0:n-1} ) p( \bm{\theta}_{t_{n-1}:t_n} | \bm{\theta}_{t_0:t_{n-1}} )  \notag \\  
    & \qquad \times p(\bm{\theta}_{t_0:t_{n-1}}|\mathbf{m}_{0:n-1}),  
\end{align}
with ${p}(\bm{\theta}_{t_0:t_{n-1}}|\mathbf{m}_{0:n-1}) \approx \sum^{N_p}_{i=1} \omega^{(i)}_{n-1} \delta_{\bm{\theta}_{t_0:t_{n-1}}^{(i)}} (\bm{\theta}_{t_0:t_{n-1}})$.
Sampling from \eqref{eqn:target_density_rbvrpf} relies on the use of an importance distribution $q(\bm{\theta}_{t_0:t_n}|\mathbf{m}_{0:n})$ that has the general form
\begin{align}  \label{eqn:general_joint_proposal_vrpf}
    q(\bm{\theta}_{t_0:t_n} | \mathbf{m}_{0:n})
     & =q(\bm{\theta}_{t_{n-1}:t_n}|\bm{\theta}_{t_0:t_{n-1} }, \mathbf{m}_{0:n}) \notag \\ 
     & \qquad \qquad \qquad \times q(\bm{\theta}_{t_0:t_{n-1}}|\mathbf{m}_{0:n}).
\end{align}
The past state trajectory $\bm{\theta}^{(i)}_{t_0:t_{n-1}}$ is drawn from a discrete distribution as per 
\begin{align} 
    q(\bm{\theta}_{t_0:t_{n-1}}|\mathbf{m}_{0:n}) = \sum^{N_p}_{i=1} v^{(i)}_{n-1} \delta_{\bm{\theta}_{t_0:t_{n-1}}^{(i)}} (\bm{\theta}_{t_0:t_{n-1}}), \notag
\end{align}
where the non-negative selection weights $\{ v^{(i)}_{n-1} \}_{1 \leq i \leq N_p}$ sums to one and the state sequence $\bm{\theta}^{(i)}_{t_{n-1}:t_{n}}$ arriving between $t_{n-1}$ and $t_n$ is sampled from $q(\bm{\theta}_{t_{n-1}:t_n}|\bm{\theta}_{t_0:t_{n-1} }, \mathbf{m}_{0:n})$. Denoting by $\{ \bm{\theta}_{t_0:t_{n}}^{(i)} \}_{1 \leq i \leq N_p}$ the $N_p$ samples drawn from the joint proposal in \eqref{eqn:general_joint_proposal_vrpf} at $t_n$, the VRPF weights update is
\begin{align}
    \widetilde{\omega}^{(i)}_n 
    & = 
    \frac{ p(\bm{\theta}^{(i)}_{t_0:t_n}|\mathbf{m}_{0:n})  }{ q(\bm{\theta}^{(i)}_{t_0:t_n} | \mathbf{m}_{0:n}) } \notag \\
    & \propto \frac{ \omega_{n-1}^{(i)} }{ v_{n-1}^{(i)} }
    \times
    \frac{p(\mathbf{m}_n | \bm{\theta}^{(i)}_{t_0:t_{n}}, \mathbf{m}_{0:n-1} ) p( \bm{\theta}^{(i)}_{t_{n-1}:t_n} | \bm{\theta}^{(i)}_{t_0:t_{n-1}} ) }{q(\bm{\theta}^{(i)}_{t_{n-1}:t_n}|\bm{\theta}^{(i)}_{t_0:t_{n-1} }, \mathbf{m}_{0:n})}, \notag
\end{align}
where $\widetilde{\omega}^{(i)}_n$ is the unnormalised particle weight for the $i$-th particle. In this paper, we chose to use the prior distribution \eqref{eqn:variable_rate_state_transition_density} as the importance distribution for $\{ \bm{\theta}_{t_{n-1}:t_{n}}^{(i)} \}_{1 \leq i \leq N_p}$, that is, $q(\bm{\theta}_{t_{n-1}:t_n}|\bm{\theta}_{t_0:t_{n-1} }, \mathbf{m}_{0:n}) = p( \bm{\theta}_{t_{n-1}:t_n} | \bm{\theta}_{t_0:t_{n-1}} )$. This leads to a simplified weighting function
\begin{align} 
    \label{eqn:weight_update_rbvrpf}
    \widetilde{\omega}^{(i)}_n
    \propto \frac{ \omega_{n-1}^{(i)} }{ v_{n-1}^{(i)} } \times p(\mathbf{m}_n | \bm{\theta}^{(i)}_{t_0:t_{n}}, \mathbf{m}_{0:n-1} ).
\end{align}
In order to mitigate the particle degeneracy problem, we apply a popular adaptive selection (or re-sampling) strategy where the particles are re-sampled according to the selection weights if the effective sample size (ESS) is lower than a pre-determined threshold or the particle set remains unaltered if the ESS is higher than the threshold. In the former case we have $v^{(i)}_{n-1} = \omega^{(i)}_{n-1}$ and $v^{(i)}_{n-1} = 1/N_p$ for the latter. Although not examined here, it is possible to use the incoming and/or future measurements to better guide the selection process. These more elaborated ``lookahead'' strategies have been successfully implemented in the context of particle filtering, see \cite{Lin2013,godsill07}.

To calculate the particle weight from \eqref{eqn:weight_update_rbvrpf}, we need to know the predictive likelihood term $p(\mathbf{m}_n | \bm{\theta}^{(i)}_{t_0:t_{n}}, \mathbf{m}_{0:n-1} )$. Next, we show that this quantity can be obtained as a byproduct when updating the Kalman filter associated with the $i$-th particle.

\subsubsection{Kalman filtering}
The Gaussian mixture approximation of the filtering distribution of the linear component requires the mean and covariance for each Gaussian density $p(\mathbf{s}_n | \bm{\theta}_{t_0:t_n}^{(i)}, \mathbf{m}_{0:n})$ as in \eqref{eqn:linear_posterior_gauss_mixture}. These two quantities can be computed using a bank of Kalman filters \cite{anderson1979}. Here we detail the calculation steps with a focus on the state transition density of the piecewise constant model. The resultant is also applicable to the jump-diffusion model and fast manoeuvring target models. Similar operations can be readily applied to the multi-hypothesis model with the transition density \eqref{eqn:vl_multiHypo_overall_density} (extended to the multiple jumps scenario). Provided the posterior mean and covariance obtained from the last time step, the predictive distribution of the $i$-th Kalman filter can be computed as per
\begin{align} \label{eqn:kalman_prediction}
    p(\mathbf{s}_n | \bm{\theta}_{t_0:t_n}^{(i)}, \mathbf{m}_{0:n-1})
    = \mathcal{N}(\mathbf{s}_n| \mu^{(i)}_{n|n-1}, \Sigma^{(i)}_{n|n-1} ),
\end{align} 
with 
\begin{align*}
    \mu_{n|n-1}^{(i)} & = e^{A\delta t} \mu_{n-1|n-1}^{(i)} + \sum_{k: \tau_k \in (t_{n-1}, t_n]} \mu_J  e^{A(t_n-\tau_k)} \mathbf{h}_J,  \\
    \Sigma_{n|n-1}^{(i)} & = e^{A \delta t} \Sigma^{(i)}_{n-1|n-1} (e^{A \delta t})^T + \Sigma_{n:n-1}.
\end{align*} 
Note $\delta t = t_n-t_{n-1}$ and $\Sigma_{n:n-1}$ is defined in \eqref{eqn:vl_conditional_pc_density}. When the new measurement $\mathbf{m}_n$ becomes available the posterior mean and covariance can be updated via the Kalman correction step, 
\begin{align} \label{eqn:kalman_correction}
\begin{split}
    \mu_{n|n}^{(i)} & = \mu_{n|n-1}^{(i)} + K_n (\mathbf{m}_n - H \mu_{n|n-1}^{(i)}  ), \\
    \Sigma_{n|n}^{(i)} & = (I - K_n H) \Sigma_{n|n-1}^{(i)}, \\
    K_n & = \Sigma_{n|n-1}^{(i)} H^T ( H \Sigma_{n|n-1}^{(i)} H^T + R)^{-1}. 
\end{split}
\end{align}
   
The predictive likelihood term needed for the computation of the importance weight can now be calculated using the predictive error decomposition \cite{harvey1990} 
\begin{align}\label{eqn:predictivelikelihood}
    &p(\mathbf{m}_n | \bm{\theta}^{(i)}_{t_0:t_{n}}, \mathbf{m}_{0:n-1} )\notag \\ 
    & = \int p(\mathbf{m}_n | \mathbf{s}_n ) p(\mathbf{s}_n | \bm{\theta}_{t_0:t_n}^{(i)}, \mathbf{m}_{0:n-1}) d \mathbf{s}_n \notag \\
    & = \mathcal{N}(\mathbf{m}_n | \mu_{\mathbf{m}_n}^{(i)}, \Sigma_{\mathbf{m}_n}^{(i)} ),
\end{align}
with
\begin{align*}
    \mu_{\mathbf{m}_n}^{(i)} = H \mu^{(i)}_{n|n-1} , \;\;\;
    \Sigma_{\mathbf{m}_n}^{(i)} = H \Sigma^{(i)}_{n|n-1} H^T + R.
\end{align*}

\subsection{Intentionality Estimation} \label{sec:intentprediction}
We now illustrate how the target latent intent may be determined from the inferred
 posterior using \eqref{eqn:linear_posterior_gauss_mixture}-\eqref{eqn:predictivelikelihood} for all introduced models in this paper, except the basic one in Section \ref{ssec:vl_basic_des_model}. The intermediate (i.e. waypoint) or final destination can be spatial point or extended regions, which can  chosen impromptu (e.g. by the non-cooperative surveillance system operator) anytime and anywhere within the state space (e.g. in the analysed scene). 

At time $t_n$ destination $\mathbf{r}_n$ is a sub-state of $\mathbf{s}_n = [ \mathbf{x}_n, \mathbf{r}_n]^{T}$ and the intent marginal posterior can be readily approximated as the following Gaussian mixture
\begin{align}
\label{eqn:r_n_gauss_mix}
    p(\mathbf{r}_n | \mathbf{m}_{0:n}) 
    \approx \sum^{N_p}_{i=1} \omega^{(i)}_n \int \mathcal{N} (\mathbf{s}_n | \mu_{n|n}^{(i)}, \Sigma_{n|n}^{(i)} ) d \mathbf{x}_n
\end{align}
We can obtain an estimate of the probability of some destination located at the spatial point $\mathbf{p}_{\mathrm{Des}}$  by evaluating $\hat{p}(\mathbf{r}_n=\mathbf{p}_{\mathrm{Des}} | \mathbf{m}_{0:n} )$. When several positions are concurrently investigated, the relative or normalised values of the produced probabilities can be compared to make a decision on the most likely intent.
 
For an extended spatial region $\mathcal{A}$ of interest, we capitalise on the fact that the introduced formulation provides the destination posterior density $p(\mathbf{r}_n| \mathbf{m}_{0:n} )$ and calculate the integral 
\begin{equation} \label{eqn:area_prob_integration}
    \mathrm{Pr}(\mathrm{Des}=\mathcal{A}|\mathbf{m}_{0:n})=\int_{\mathcal{A}} p(\mathbf{r}_n|\mathbf{m}_{0:n})d \mathbf{r}_n, 
\end{equation} 
which can be easily evaluated for a Gaussian (or a mixture of Gaussians) posterior density; regions of rectangular (or cuboid) shape can also simplify the calculations (e.g. using the multivariate cumulative distribution functions). For other distributions and geometric shapes, approximations can be applied. 
This will be treated in more detail in future work and it is not expected to lead to drastically different outcomes to approximating the area of interest with $k$ simpler shapes (e.g. rectangles, ellipsoids, etc.) via $\sum_k\int_{\mathcal{A}_k} p(\mathbf{r}_n|\mathbf{m}_{0:n})d \mathbf{r}_n$. Similarly, a number of spatial areas can be routinely treated by computing \eqref{eqn:area_prob_integration} for each one independently.

\section{Results} \label{sec:Results}
The performance of the proposed jump particle filtering for the joint estimation of a target's state and intent is assessed in this section. First, we use synthetically generated data to evaluate all models in Section \ref{sec:ProposedModels} prior to evaluating the performance of selected suitable ones on real radar measurements of a small drone. Reasonable values for the (hyper-)parameters of all of the considered augmented state models are chosen below based on their physical interpretation (e.g. reversion constants, drag coefficient, etc.) for both synthetic and real data, without extensive manual fine-tuning. It is possible to learn these  based on maximising the (marginal) likelihood from a number of example trajectories. Within the Markovian set-up here,  the likelihoods can be directly computed based on the prediction error decomposition of the Kalman or particle filter \cite{sarkka2023bayesian}. Learning model parameters and a detailed empirical analysis of the models sensitivity to the hyperparameters choice can be investigated in future work. However, from the evaluation below and previous work on similar models \cite{ahmad16_ou, ahmad18_bd, gan_liang_ahmad_godsill_2020, liang19_pseudo, gan2020modeling, gan2021levy} optimising the model hyper-parameters, beyond selecting  sensible values (e.g. process noise standard deviations), does not substantially impact the overall inference accuracy. An exception may be the reversion parameters for the destination and minimal fine-tunning might be required for those.

\subsection{Simulated Data}
The accuracy of the estimation of the target location, velocity and destination/waypoint position (i.e. mean root mean square error RMSE of positional/velocity estimates across a trajectory) are listed in Table \ref{tab:vl_simu_rmse_non_fast_manoeuring} for 100 simulated target tracks with a nearly Constant Velocity (CV) model; its dynamical noise has standard deviation $\sigma_d = \sqrt{14}$ and $d=x,y,z$. Table \ref{tab:vl_simu_rmse_fast_manoeuring} is for 100 fast maneuvering targets generated with the jump diffusion model ($\mu_J = 0$ and $\sigma_J = 50$). Each of the trajectories has $N=3$ to $N=5$ waypoints randomly placed within the scene. A Gaussian noise with $\sigma^R_d=15$m is added to the 3-D positional data to produce the measurements (i.e. noisy target locations in Cartesian coordinates) for the meta-level tracker. Results from the following methods are reported:
\begin{itemize}
    \item \textit{VL-D-KF}: baseline model in Section \ref{ssec:vl_basic_des_model} with a Kalman Filter (KF) as in \cite{liang2021SSPD}.
    \item \textit{VL-PC-RBVRPF} for piecewise constant destination model in Section \ref{ssec:vl_piecewiseconst} with Rao-Blackwellised variable rate particle filter. 
    \item \textit{VL-PC (known $\tau$)} for model in Section \ref{ssec:vl_piecewiseconst} and KF since jump/switch times are assumed known. 
    \item \textit{VL-PC (known $\tau$ and $\mathbf{r}$)} for model in Section \ref{ssec:vl_piecewiseconst} and KF with known jump/switch times and waypoints/destination.   
    \item \textit{VL-MultHyp-RBVRPF} for the multiple hypothesis model in Appendix A and RBVRPF; the $N_{\mathcal{D}}$ nominal destinations are assigned to the true trajectory waypoints.
    \item \textit{VL-JD-RBVRPF} for the jump diffusion model in Section \ref{ssec:vl_jumpdiff} and RBVRPF.  
     \item \textit{VL-FMT-RBVRPF} for the fast maneuvering target model in Section \ref{ssec:vl_fmt} and RBVRPF.   
\end{itemize}

\begin{table}[!t]
\centering 
\caption{Target position, target velocity and destination location estimation performance (RMSE) across 100 realisations of trajectories containing waypoints for (non-fast-maneuvering) targets.}
\label{tab:vl_simu_rmse_non_fast_manoeuring} 
\begin{tabular}{l|c|c}
\toprule\hline  
Methods & \textsl{Pos. (m) -- Vel. (m/s) Est.} 
& \textsl{Dest. Est. (m)} 
\\
\hline 
VL-D-KF     &  2.03 -- 0.33 & 19.31 \\  
\hline
VL-JD-RBVRPF & 1.74 -- 0.34 & 14.81 \\  
\hline
VL-PC-RBVRPF & 1.75 -- 0.37 & 14.13 \\  
\hline
VL-FMT-RBVRPF & 2.18 -- 0.42 & 23.92 \\  
\hline 
VL-MultHyp-RBVRPF & 1.46 -- 0.31 & \textbf{12.06} \\
\hline
VL-PC (known $\tau$) & \textbf{1.43} -- 0.29 & 12.93 \\  
\hline
VL-PC (known $\tau$ and $\mathbf{r}$) &  1.55 -- \textbf{0.24} & -- \\
\hline\bottomrule
\end{tabular}
\end{table}
\begin{table}[!t]
\centering 
\caption{Target position, target velocity and destination location estimation performance (RMSE) across 100 realisations of trajectories containing waypoints for fast-maneuvering targets.}
\label{tab:vl_simu_rmse_fast_manoeuring}
\begin{tabular}{l|c|c}
\toprule\hline  
Methods & \textsl{Pos. (m) -- Vel. (m/s) Est.} 
& \textsl{Dest. Est. (m)} 
\\
\hline 
VL-D-KF & 2.78 -- 0.55 & 26.95 \\  
\hline
VL-JD-RBVRPF & 2.54 -- 0.51 & 20.23 \\  
\hline
VL-PC-RBVRPF & 2.58 -- \textbf{0.49} & \textbf{20.12} \\  
\hline
VL-FMT-RBVRPF & \textbf{2.21} -- 0.53 & 25.53 \\  
\hline
\bottomrule
\end{tabular}
\end{table}

Table \ref{tab:vl_simu_rmse_non_fast_manoeuring} shows that the basic VL-D-KF model achieves the poorest estimation performance compared to the other proposed approaches in terms of positional estimates. Without known destination switch times, the multi-hypothesis model delivers the best performance followed by the more general VL piece-wise constant model. However, in numerous scenarios (e.g. surveillance of non-cooperative targets such as drones) it is not possible in practice to have prior information on the target possible waypoints or nominal destinations as with the multi-hypothesis model. Given the underlying CV motion model of the simulated data, using more sophisticated movement models aimed at targets undertaking fast maneuvers can degrade the overall tracking performance (e.g. compared to models using CV) as seen in Table \ref{tab:vl_simu_rmse_non_fast_manoeuring}. Conversely, the benefits of such advanced modelling is visible in the results in Table \ref{tab:vl_simu_rmse_fast_manoeuring} where the fast maneuvering target model has the best target position estimates, whereas the VL-PC has the most accurate destination/waypoints inference results. This can be attributed to the ability of the VL-PC to effectively capture dynamic changes in the object intent. On the other hand, the VL-D-KF for Gaussian and linear settings, as expected, depicts the least accurate estimates of both target and destination/waypoints locations in such scenarios. 

For velocity, VL-PC delivers the most accurate average estimates from both tables. FMT model has the highest errors with the CV-model-based data, which is expected. Nonetheless, augmenting the target kinematic state with its latent intent generally enables more accurate velocity estimation, albeit marginally.  A more detailed analysis of the velocity estimates (e.g. as a function of distance to the sought waypoints or final destination) can be explored in future work. This includes incorporating higher order kinematics (such as velocity) in the hidden destination dynamical model.
\subsection{Real Radar Data for Intent and Threat Detection}
\begin{figure}[t] 
\centering   
\includegraphics[width=1\linewidth]{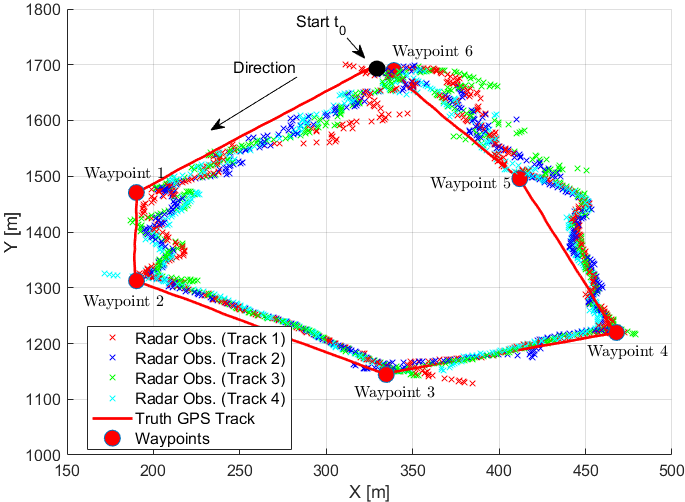}  
\caption{Radar observations of a drone surveying an area with predefined waypoints; truth trajectory from onboard GPS is shown. }
\label{fig:real_drone_tracks_profile_1}
\end{figure}

We now employ radar observations from Thales Gamekeeper 16U staring radar system. This sensor has been specifically designed for high performance detection and automatic recognition of small drones, which can be slow and fly at low altitudes, within a $7.5$km range in its current configuration. The data was collected during the SESAR SAFIR live demonstrations near Port of Antwerp, Belgium, see \cite{jahangir2020robust}. It contains the 3-D Cartesian coordinates\footnote{Due to the sensitive, commercial, nature of the radar raw detections, they could not be discussed in this paper. Instead, an \textit{experimental} processing chain, including clustering, nearest neighbor data association and basic extended Kalman filter, is applied by the sensor and its output is used for evaluating the proposed algorithms.} of targets of interest within the radar field of coverage, specifically: a) a DJI Inspire II quadcopter drone (diameter $\approx0.5$m and weight $\approx3.5$ kg) undertaking a site surveying task within an authorised flying zone $\mathcal{A}$ whilst following waypoints-driven trajectories as in Figure \ref{fig:real_drone_tracks_profile_1}; and b) unidentified \textit{opportune} airborne objects (e.g. microlights, etc.). 

Here, the baseline VL-D-KF and the introduced VL-PC-RBVRPF (unknown $\tau$ and $\mathbf{r}$) techniques are selected for testing where four different performance aspects are examined, namely accuracy of target tracking with and without intent estimation, ability to predict the target next destination/waypoint, revealing malicious intent and early threat detection. Models for fast maneuvering targets are not considered since the drone waypoints-guided tracks are relatively smooth, see Figure \ref{fig:real_drone_tracks_profile_1}.

\subsubsection{Intent-aware tracking} 
\begin{figure}[!t] 
\centering   
\includegraphics[width=0.95\linewidth]{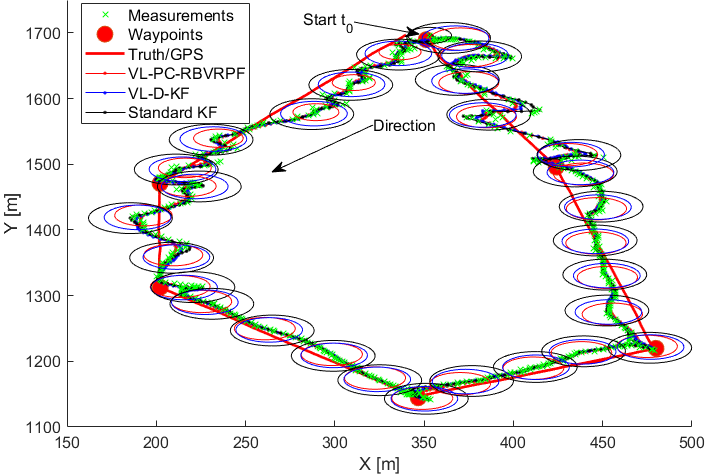}  
\caption{Drone track estimation results (green crosses are radar observations). The dotted solid lines are posterior mean filtering estimates for VL-D-KF, VL-PC-RBVRPF and standard Kalman filter (i.e. no intent modelling); posterior confidence ellipses are shown.}
\label{fig:real_drone_tracks_profile_1_target_traj}
\end{figure}

The kinematic state estimation results are displayed in Figure \ref{fig:real_drone_tracks_profile_1_target_traj} for the measurements of Track 2 in Figure \ref{fig:real_drone_tracks_profile_1} where ellipsoids represent the 95$\%$-confidence region of the estimates; the outcome from a standard Kalman filter (i.e. without intent modelling) is also shown. It can be noticed that both VL-D-KF and VL-PC-RBVRPF intent modelling lead to more certain estimates with smaller ellipsoids compared to the standard KF, especially near the waypoints. Their tracks are also slightly closer to the ground-truth. The mean RMSE of the standard KF, VL-D-KF and VL-PC-RBVRPF for all four trajectories in Figure \ref{fig:real_drone_tracks_profile_1} are $13.4$m, $12.9$m and $12.6$m, respectively. Whilst VL-PC-RBVRPF has the lowest RMSE of positional estimates and smaller ellipsoids, there are a few occasions where the ground-truth falls outside its confidence ellipsoids. This can be due to a possible misalignment between the ground-truth data from the GPS-based navigation system onboard the drone  and the radar measurements, namely due to inaccuracies in the assumed sensor true orientation-tilt. This is  visible in Figure \ref{fig:real_drone_tracks_profile_1} (e.g. before waypoint 1). We recall that the available real data has already been processed/filtered, which limits the ability to achieve bigger improvements in the tracking performance.
\subsubsection{Waypoints learning} 
\begin{figure}[t]
    \centering
    \begin{subfigure}[t]{1\linewidth} 
    \centering
    \includegraphics[width=1\linewidth]{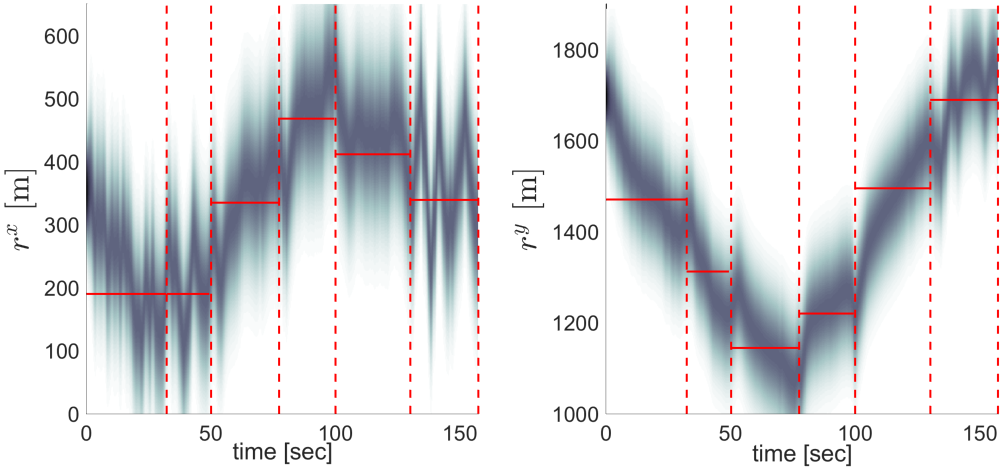}
    \caption{VL-D-KF} 
    \label{subfig:real_drone_tracks_profile_1_waypoint_VL_D}
    \end{subfigure} 
    \begin{subfigure}[t]{1\linewidth} 
        \centering
        \includegraphics[width=1\linewidth]{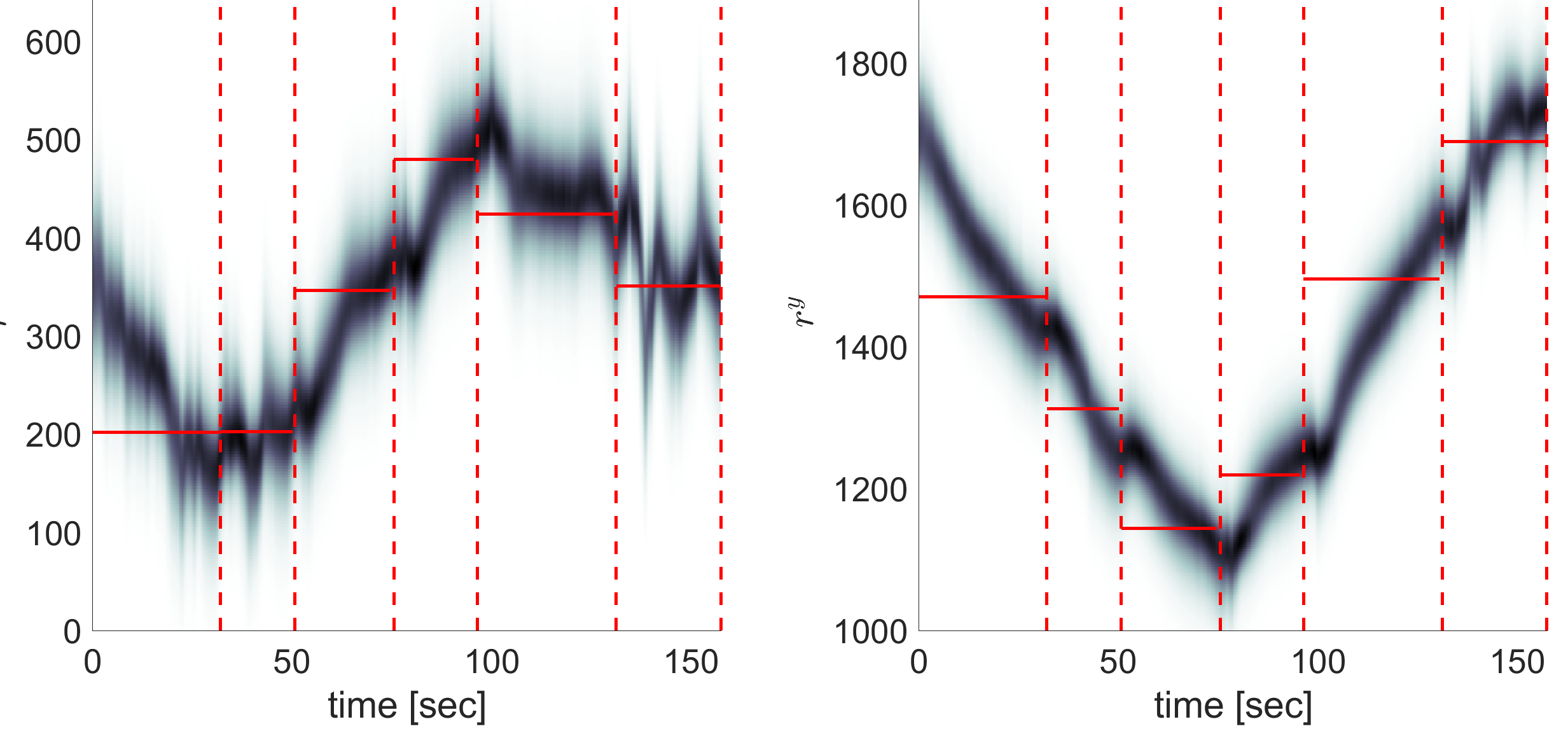}
        \caption{VL-PC-RBVRPF} \label{subfig:real_drone_tracks_profile_1_waypoint_VL_RBVRPF_trimmed}
    \end{subfigure} 
    \caption{Waypoint estimates showing filtering posterior of the destination location along the X and Y axes. Vertical red dashed lines indicate time instants when the drone reached a waypoint. Red solid horizontal lines are true waypoint positions in the corresponding axis.}
    \label{fig:real_drone_tracks_profile_1_waypoint_VL_PC}
\end{figure}

\begin{figure}[t] 
\centering   
\includegraphics[width=1\linewidth]{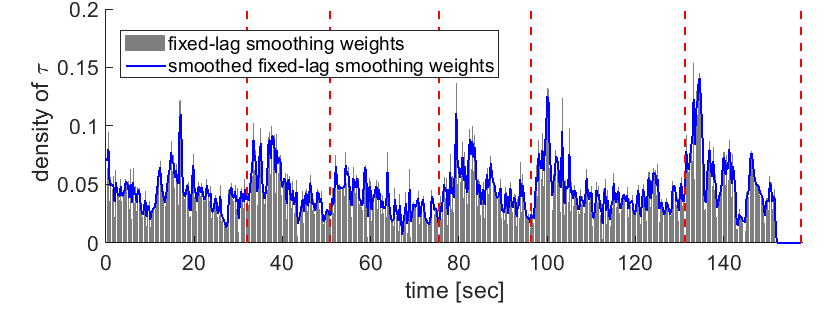}  
\caption{Destination switching time $\tau$ with smoothing; vertical red dashed lines indicate time instant drone reached a waypoint.}
\label{fig:dest_switiching_time_detection}
\end{figure}

Figure \ref{fig:real_drone_tracks_profile_1_waypoint_VL_PC} shows waypoints estimation results from VL-PC-RBVRPF and VL-D-KF on the same drone trajectory in Figure \ref{fig:real_drone_tracks_profile_1_target_traj}.  It depicts the inferred posterior for the $r^{x}_{n}$ and $r^{y}_{n}$, denoting the latent destination X and Y position, respectively. Both are members of the inferred augmented system state $\mathbf{s}_n$. 
The figure demonstrates the ability of the introduced VL-D-KF and VL-PC-RBVRPF to predict the waypoints well in advance. The former exhibits higher variability in predictions (i.e. more spread posteriors) compared with VL-PC-RBVRPF which delivers higher response to change in (intermediate) destinations and more certain estimates of the waypoint location. Figure \ref{fig:dest_switiching_time_detection} displays the VL-PC-RBVRPF destination switching time detection with  more jumps visible right after the target aims for a new waypoint (i.e. vertical dash red lines marking time instant when the drone reached a waypoint), indicating the ability of the VL-PC-RBVRPF algorithm to quickly identify new intents. It is noted that beyond improved tracking results, early prediction of the target next waypoint can enable more effective deployment of countermeasures and proportionate mitigation strategies (e.g. against malicious drone activities). 


\subsubsection{Drone to leave/enter an authorised/prohibited region $\mathcal{A}$} 
\begin{figure}[t] 
\centering   
\includegraphics[width=\linewidth]{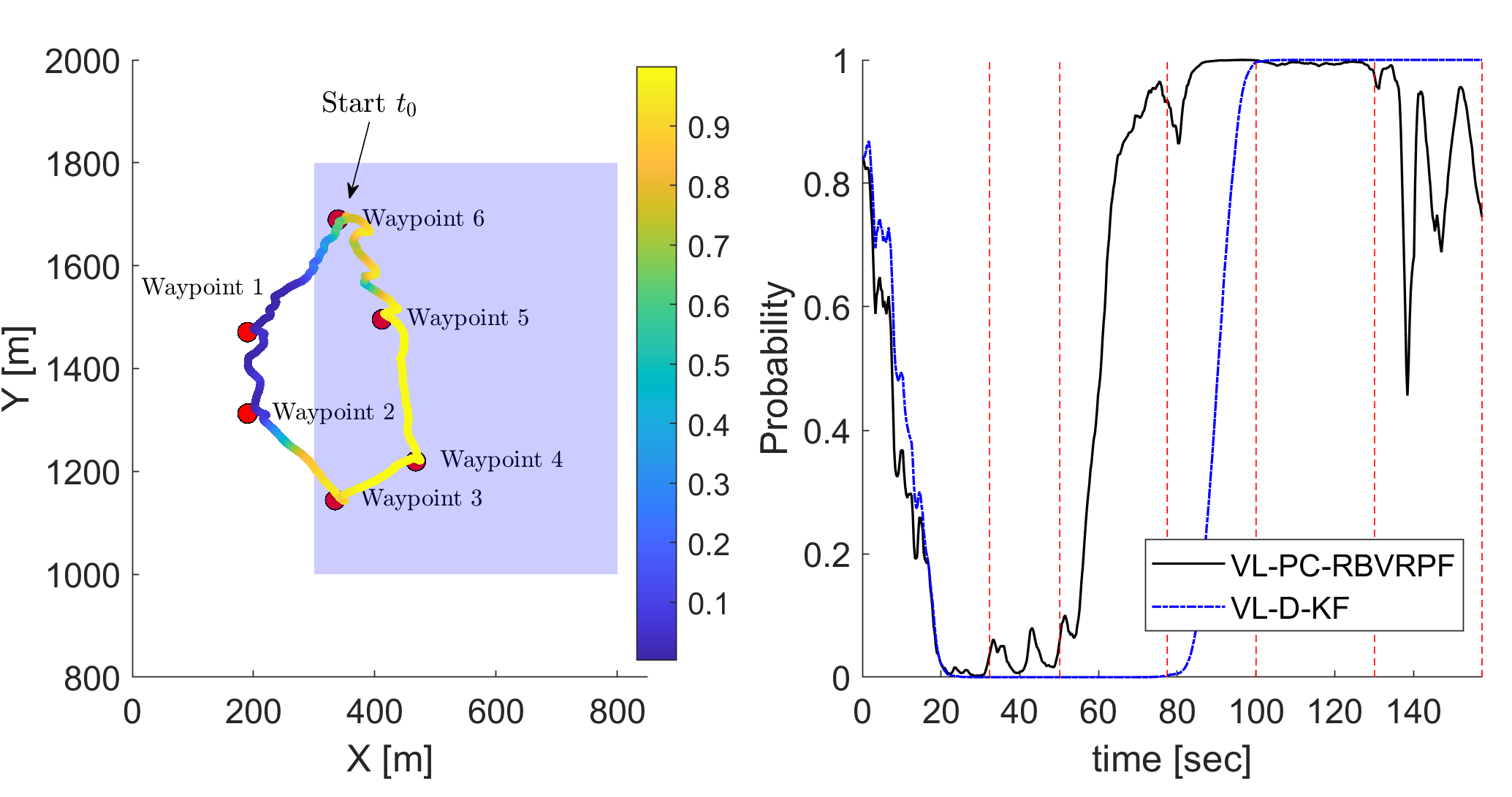}  
\caption{Threat detection with the drone exiting and re-entering a permitted zone $\mathcal{A}$. Left:  target trajectory, waypoints and probability $\mathrm{Pr}(\mathrm{Des}=\mathcal{A}|\mathbf{m}_{0:n})$  as provided by VL-PC-RBVRPF; $t_0$ is the flight start time. Right: probability of destination being within the zone given by the two the proposed algorithms VL-D-KF and VL-PC-RBVRPF; vertical dashed lines are time instant the drone reach each of the waypoints.}
\label{fig:antwerp_area_threat_safe_zone_exp2}
\end{figure}

This uses observations of a track from Figure \ref{fig:real_drone_tracks_profile_1} as in \cite{liang2021SSPD} and the results are depicted in Figure \ref{fig:antwerp_area_threat_safe_zone_exp2}. They illustrate the ability of the piece-wise constant destination model to respond quicker to change in intent compared with the basic model. This is most visible when the sUAS re-enters $\mathcal{A}$ between waypoints 2 and 3. The VL-PC-RBVRPF predicted re-entry early (i.e. after the target departed waypoint 2) and reacts to the UAS reaching waypoint 6 near the boarder of $\mathcal{A}$.
\begin{figure}[b] 
\centering   
\includegraphics[width=1\linewidth]{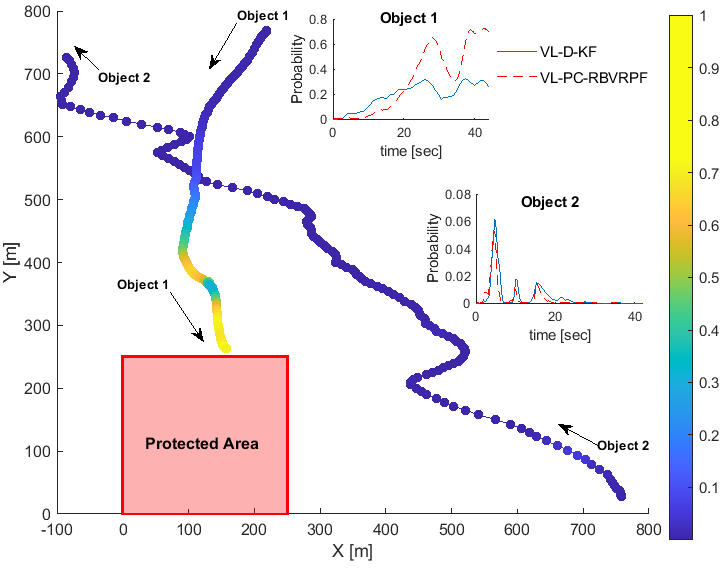}  
\caption{Threat detection example with results of estimating the probability that the protected region is the intended destination of two targets of interest. The trajectories color reflects the probabilities produced by the VL-PC-RBVRPF method. Subplots show the calculated probabilities over-time for VL-PC-RBVRPF and VL-D-KF.}
\label{fig:antwerp_area_threat_non_drone_examples}
\end{figure}  
\subsubsection{Malicious intent to reach a restricted region $\mathcal{A}$} The aim is to reveal, as early as possible, if either of the two tracked targets in Figure \ref{fig:antwerp_area_threat_non_drone_examples} intend to enter a protected spatial area; $\mathcal{A}$ is (artificially) set near the radar. It can be seen both algorithms correctly determined that object 2 was not posing a threat. Whereas the probability that $\mathcal{A}$ is the destination Object 1 increased dramatically as the target approaches and/or heads toward the protected area. The VL-PC-RBVRPF responded quicker to a potential change of intent (including dip at time instant $t_n\approx35$sec compared to VL-D-KF where its $\mathrm{Pr}(\mathrm{Des}=\mathcal{A}|\mathbf{m}_{0:n})$ exceeds 50\% up to 20-30 seconds before Object 1 enters $\mathcal{A}$. This demonstrates the ability of the proposed intent inference approach to provide tactical advantage in terms of early detection of threats and provide threat prioritisation capability (e.g. system operator not distracted by Object 2).      

Finally, the presented Bayesian approach provides the means to continuously estimate the potentially dynamically changing intent (i.e. waypoints and final destination) of targets and leverage this information to deliver improved tracking results (i.e. state estimation). It seamlessly and automatically incorporates salient behavioral attributes (such as distance to destination, heading, velocity or higher order kinematics) indicative of the object's future actions as per the target goal-driven motion model as seen in Figures \ref{fig:antwerp_area_threat_safe_zone_exp2} and \ref{fig:antwerp_area_threat_non_drone_examples}. On the other hand, there are numerous threat or intent assessment schemes in the surveillance domain that rely on specific features from the target radar signature, e.g. micro-Doppler characteristics as in \cite{de2025emerging}, and/or hand-crafted attributes from the estimated target motion, such as the shape or frequency of the followed trajectory, acceleration profile, proximity to an asset, heading angle and other; see \cite{yun2023estimation} for a recent overview. Whilst results from both methodologies can be combined at the command and control (C2) system level, performance comparisons require specialised data collection campaigns advised by end-user operational requirements (e.g. for counter drone applications for fixed-site protecting such as airports). This is outside the scope of this work.
\section{Conclusion}\label{sec:conclusion}
In this paper, we propose and illustrate the potential of a novel Bayesian meta-level tracker for the joint estimation of an object latent kinematic state and destination/waypoints from noisy sensory observations. It: a) can handle dynamically changing intents; b) is in principle agnostic to both the sensing technology and target type; c) is based on a particle filtering algorithm for sequential inference with minimal training data requirements (similar to classical well-understood object trackers); and d) utilises continuous-time models that naturally permit treating irregular and asynchronous sensory measurements, including from multiple sensors. Future work will extend the introduced framework to address multi-target scenarios, such as of coordinated groups of objects, and the data association aspect.

\appendices
\section{Multi-Hypothesis Latent Intent Model} 
For $N_{\mathcal{D}}$ predefined destinations (e.g. spatial points or regions), we have $N_{\mathcal{D}}+1$ hypotheses inclusive of the null hypothesis $\mathcal{H}_0$. We recall that $\mathcal{H}_0$ signifies that the target intent is none of these nominal endpoints. This multi-hypothesis intent model follow from that in Section \ref{ssec:vl_piecewiseconst} with an additional indicator parameter $c_k$. 
 Assume for simplicity that there is only one jump that arrives at $\tau_k$ within the time interval $[t_{n-1},t_{n}]$; $\tau^+_k$ and $\tau^-_k$ denote the timings right after and right before the jump time $\tau_k$, respectively. The transition process of $\mathbf{s}=[x,\dot{x},r^x]^T$ from $t_{n-1}$ to $t_{n}$ can be divided into the three phases,
\begin{enumerate}
    \item {From $t_{n-1}$ to $\tau_k^-$:}  
    \begin{equation} \label{eqn:trans_before_tau_k}
        \hspace*{-2em}
        p(\mathbf{s}(\tau_k^-)|\mathbf{s}_{n-1})=\mathcal{N} (\mathbf{s}(\tau_k^-) | F_{\tau_k-t_{n-1}} \mathbf{s}_{n-1}, \widetilde{Q}_{\tau_k-t_{n-1}} ),
    \end{equation}
    where $\widetilde{Q}_{\Delta t} = \int^{\Delta t}_0 e^{A(\Delta t-u)} \mathbf{h}_B \mathbf{h}_B^T {e^{A(\Delta t-u)}}^T du$ and $F_{(\cdot)}$ is defined in \eqref{eqn:expm(A)_basic}. 
    \item {From $\tau_k^-$ to $\tau_k^+$:}
    \begin{align} \label{eqn:trans_tau_k}
        & p(\mathbf{s}(\tau_k^+)|\mathbf{s}(\tau_k^-)) =  \\ 
        & \qquad \qquad \mathcal{N}(\mathbf{s}(\tau_k^+)|F^{\pm}(c^+_k)\mathbf{s}(\tau_k^-)+M^{\pm}(c^+_k), Q^{\pm}(c^+_k)), \notag 
    \end{align} 
    where $c^+_k$ is the indicator value right after $\tau_k$. Note here that $c^+_k=c_k$ and $c^-_k=c_{k-1}$.
    \item {From $\tau_k^+$ to $t_{n}$:}
    \begin{equation} \label{eqn:trans_after_tau_k}
        p(\mathbf{s}_{n}|\mathbf{s}(\tau_k^+))=\mathcal{N} (\mathbf{s}_{n} | F_{t_{n}-\tau_k} \mathbf{s}(\tau_k^+), \widetilde{Q}_{t_{n}-\tau_k} ).
    \end{equation}
\end{enumerate}  
In the first and the last phases, the transition and covariance matrices take the same functional forms since during these two periods the latent destination remains constant. 

At $\tau_k$, there is an instant state transition taking place under the guidance of a new indicator value $c^+_k$. The exact forms of the matrices in the Gaussian density function in \eqref{eqn:trans_tau_k} depend on the type of the corresponding hypothesis. Define $c_k \in \{ 0, 1, \ldots, N_{\mathcal{D}} \}$ and then we have
\begin{align*}
    c_k = 
    \begin{cases}
      j, & j \in \{ 1, \ldots, N_{\mathcal{D}} \} \\ 
      0, & \text{null hypothesis}
    \end{cases}
\end{align*}
In the case of the $j^{\textrm{th}}$ alternative hypothesis (i.e. the $j^{\textrm{th}}$ predefined destination), the matrices are given by
\begin{align*}
    F^{\pm} (c^+_k=j) = 
    \begin{bmatrix}
    I_{2 \times 2} & 0_{2 \times 1} \\
    0_{1 \times 2} & 0 
    \end{bmatrix},
    M^{\pm} (c^+_k=j) = [0, 0, p^x_j]^T
\end{align*}
and $Q^{\pm} (c^+_k=j)  = \textrm{diag}([0, 0, (\sigma^x_j)^2])$; $I_{a \times b}$ denotes a $a$-by-$b$ identity matrix and similarly $0_{a \times b}$ is a $a$-by-$b$ zero matrix. Here, $p_j^x$ is the position of the $j^{\textrm{th}}$ destination along $x$ axis while $\sigma_j^x$ describes the extent-orientation of this endpoint (e.g. the shape and orientation of an ellipse in a 2/3-D Cartesian coordinate system). For a spatial point location, $\sigma_j^x=0$ and the Gaussian in \eqref{eqn:trans_tau_k} collapses to a Dirac delta function. On the other hand, the matrices for the null hypothesis ($j=0$) are
\begin{align*}
    F^{\pm}(c_k^+=0)=I_{3 \times 3}, \;\; 
    M^{\pm}(c_k^+=0)= \mu_J e^{t_{n}-\tau_k} \mathbf{h}_J
\end{align*}
and $Q^{\pm}(c^+_k=0)=\sigma_J^2 e^{A(t_n-\tau_k)} \mathbf{h}_J \mathbf{h}_J^T {e^{A(t_n-\tau_k)}}^T$. 

By marginalising out $\mathbf{s}(\tau_k^-)$ and $\mathbf{s}(\tau_k^+)$ in \eqref{eqn:trans_before_tau_k} through \eqref{eqn:trans_after_tau_k}, it can be shown that the target state transition density conditioned on the indicator variable and the jump time is 
\begin{align} \label{eqn:vl_multiHypo_overall_density}
    p(\mathbf{s}_n|\mathbf{s}_{n-1},c_k,\tau_k) = \mathcal{N} (\mathbf{s}_n | \breve{\mu}_{n:n-1}, \breve{\Sigma}_{n:n-1})
\end{align}
with 
\begin{align*}
    \breve{\mu}_{n:n-1} & = F_{t_n-\tau_k} \big( F^{\pm} (c^+_k) F_{\tau_k-t_n} \mathbf{s}_{n-1} + M^{\pm}(c^+_k) \big) \\
    \breve{\Sigma}_{n:n-1} & = F_{t_n-\tau_k} \big( F^{\pm} (c^+_k) \widetilde{Q}_{\tau_k-t_{n-1}} F^{\pm} (c^+_k)^T \\
    & \qquad + Q^{\pm}(c^+_k) \big) F_{t_n-\tau_k}^T + \widetilde{Q}_{t_n-\tau_k}
\end{align*}
While the density given above is with regard to a single jump occuring between $t_{n-1}$ to $t_n$, it is straightforward to generalise the result to multiple jumps. In particular, \eqref{eqn:vl_multiHypo_overall_density} will be equivalent to \eqref{eqn:vl_conditional_pc_density} if the null hypothesis is in effect at all jump times within $(t_{n-1},t_n]$. 

There are various possible choices for the prior distribution of the hypothesis indicator $\{ c \}$ and the switching times $\{ \tau \}$. For example, we adopt the following joint distribution
\begin{align} \label{eqn:tau_c_prior_multiHypo}
    p(c_k, \tau_k| c_{k-1}, \tau_{k-1})=p(c_k|c_{k-1},\tau_k)p(\tau_k|\tau_{k-1})
\end{align}
where the times between jumps are Gamma distributed as in Section \ref{ssec:vl_piecewiseconst} and the indicator variable evolves according to a fixed Markov transition probability matrix. More sophisticated priors may be devised, e.g. with the switching probabilities of the indicator dependant on the jump times $\{ \tau_k, \tau_{k-1} \}$.

Let $\bm{\theta}_k = \{ \tau_k, c_k \}$ is for the intent switch time and the indicator; their joint evolution is described by \eqref{eqn:tau_c_prior_multiHypo}. Compared to \eqref{eqn:theta_prior}, the prior for $\bm{\theta}$ in the multi-hypothesis model is
\begin{align*}
    p(\bm{\theta}_{t_0:t_n}) = S(t_n, \tau_{K_n})p(c_0)\prod_{k=1}^{K_n} p(c_k|c_{k-1},\tau_k)p(\tau_k|\tau_{k-1}),
\end{align*}
and the inference approach in Section \ref{sec:inference_method} can be applied with this definition of $\bm{\theta}_k$.

Finally, for the $j^{\text{th}}$ destination located at $p_{j}^d$ ($d=x,  y, z$ for 3-D space), which can  either a point or an extended region as per $\sigma_j^d$, and given the available sensory observations $\mathbf{m}_{0:n}$, e.g. from model in \eqref{eqn:linear_gauss_measurement_model}, the probability of each of the $N_{\mathcal{D}}+1$ hypotheses can be obtained via
\begin{align}\label{eqn:multiHypoIntent}
   \mathrm{Pr}(\mathcal{H}_j|\mathbf{m}_{0:n}) & = p(c_{K_n}=j|\mathbf{m}_{0:n}),~~~j =0,1,..., N_\mathcal{D}.
\end{align}
Since the multi-hypothesis model explicitly incorporates $\mathcal{H}_0$ and its probability can be calculated taking into account the $N_{\mathcal{D}}$ destinations, a reasonable threshold on the intent probability can used to admit one of the $N_{\mathcal{D}}+1$ hypotheses. 

\section*{Acknowledgment}
This work was funded by the Defence Science and Technology Laboratory (DSTL) via DASA grant DSTLX1000144447 and the US Army and UK Ministry of Defence via the SIGNetS project (Cooperative Agreement Number W911NF-20-2-0225). The views and conclusions contained in this paper are of the authors and should not be interpreted as representing the official policies, either expressed or implied, of the US Army Research Laboratory, the MOD, the US Government or the UK Government. 
\\Authors also thank Thales for providing the real radar data.
\bibliographystyle{IEEEtran}
\bibliography{IEEEfull,references}

%








\end{document}